# Determining the Absolute Astrometric Error in Chandra Source Catalog Positions

Short title: Absolute Astrometric Error in CSC Positions

Arnold H. Rots

Smithsonian Astrophysical Observatory,

60 Garden Street, MS 67, Cambridge, MA 02138

arots@cfa.harvard.edu

and

Tamás Budavári

Dept. of Physics and Astronomy, The Johns Hopkins University,

3400 North Charles Street, Baltimore, MD 21218

Submitted to: Astrophysical Journal Supplement Series

Received: 6 October 2010

Revised: 17 November 2010

Accepted: 17 November 2010

## Abstract

Although relative errors can readily be calculated, the absolute astrometric accuracy of the source positions in the Chandra Source Catalog (CSC), Version 1.0, is *a priori* unknown. However, the cross-match with stellar objects from the Sloan Digital Sky Survey (SDSS) offers the opportunity to compare the apparent separations of the cross-matched pairs with the formally calculated errors. The analysis of these data allowed us to derive a value of 0.16″ for the residual absolute astrometric error in CSC positions. This error will be added to the published position errors in the CSC from now on, starting with CSC, Version 1.1.

# 1 Introduction

The source positions in the Chandra Source Catalog (CSC; Evans et al., 2010) are characterized by error ellipses (circles for Version 1 of the CSC) which are based on the spatial distribution of the photons in the individual source detections. In the case of multiple detections of the same source in different observations the error ellipse of that source is derived from the error ellipses associated with the individual detections. These error ellipses provide a good measure of the statistical uncertainty of the location of the source in the frame of the observation, but leave out a series of potential sources of error that are external to the observation:

- The error in the mean aspect solution for the observation; clearly, the effect of this error will be diminished when multiple detections of the same source are combined.
- The calibration of the geometry of the spacecraft, in particular the optical axes of the aspect camera and the High Resolution Mirror Assembly (HRMA).
- The astrometric errors in the Guide Star Catalog; this should be very small.
- The calibration of the geometry of the focal plane, its projection on the detectors, and the distortions therein.

For all practical purposes, we shall combine these errors and call it an astrometric systematic error, even though not all of its components are truly systematic. The intent of this study is to derive the value of this compound quantity in order to add it to the CSC statistical position error so as to obtain a reliable absolute error for each of the CSC sources.

Using the Sloan Digital Sky Survey (SDSS; York et al., 2000) object catalog (Abazajian et al., 2009) we have the opportunity to compare the formal statistical errors with the measured separations of CSC-SDSS cross-match pairs. A statistical analysis of these data allows us to determine an accurate value for the astrometric systematic error.

This study is part of a larger project characterizing the contents of the CSC that will be described in more detail in a forthcoming publication (Primini, et al., 2010).

## 2. Cross-identification

We use the probabilistic algorithm of Budavári & Szalay (2008) to cross-match the CSC with the Seventh Data Release of the Sloan Digital Sky Survey (SDSS DR7; Abazajian et al., 2009). Using Bayesian hypothesis testing, one can objectively determine the quality of an association, which depends only on the measured positions and the astrometric uncertainties of the given sources. The Bayes factor is computed for every possible candidate association using a constant $\sigma_s = 0.1''$ uncertainty for the SDSS sources and a varying $\sigma_x$ positional error for each Chandra detection that is determined from the 95% accuracy limit $\varepsilon_0$. In the high-precision regime, the dimensionless Bayes factor is calculated as

$$B_{xs} = \frac{2}{\sigma_x^2 + \sigma_s^2} \exp\left\{-\frac{\psi_{xs}^2}{2(\sigma_x^2 + \sigma_s^2)}\right\}$$

where the uncertainties, as well as the angular separation $\psi_{xs}^2$ between the two sources, are measured in radians. Thresholding on the above formula is not equivalent to cutting on the angular separation because of the varying uncertainties in Chandra. This has proven to be superior for cross-matching GALEX and SDSS sources (Heinis et al., 2009) that exhibit a similar behavior.

Next we assign probabilities to the candidates based on a uniform prior that is determined from the ensemble statistics of the matched catalog in a self-consistent way, as described in Budavári & Szalay (2008) and Heinis et al. (2009). One may think of the prior as the average probability that a given source pair represents a physical match and it is expressed as

$$P = \frac{N_*}{N_x.\,N_s}$$

where $N_x$ and $N_s$ are the number of sources from either catalog in the intersection of the coverage of the two catalogs and $N_*$ the number of true cross-match pairs, all three scaled to the entire sky.

The posterior for each association is then given by

$$P_{xs} = \left[1 + \frac{1-P}{B_{xs}P}\right]^{-1}$$

that is reported in the cross-match catalog for each associations in addition to the angular separations and the Bayes factors. We note that, for a uniform $P$ prior, a threshold on the posterior translates directly into a Bayes factor cut. However, the interpretation of the probability is much more straightforward. Consistency requires that $N_*$ is equivalent to the sum of the posteriors $P_{xs}$ over all source pairs. Most catalogs in the subsequent analysis use a $P_{xs} > 0.9$. The CXC-SDSS cross-match catalog (version 1.0) is an exception, probably because a number of probabilities are underestimated due to the missing astrometric error that is the subject of this paper. However, we will apply the requirement $P_{xs} > 0.9$ for the purpose of this study.

## 3 Procedure

The CSC-SDSS cross-match catalog contains 7989 objects that are classified as stars in the SDSS catalog. Since these sources are, by their nature, point-like we assume their optical and X-ray positions to be well-determined and coincident. We have further narrowed the sample down by requiring the match probability to be greater than 90%. The resultant sample contains 6310 CSC-SDSS object pairs which are uniquely associated with 9476 source detections in individual observations; these 9476 objects were used for this analysis. By using the combined (CSC-SDSS) spatial error estimate of each object pair as the independent variable and analyzing the statistical distribution of the measured separations, it is possible to derive the value of the missing absolute astrometric error in the CSC. The assumption here is that the astrometric error is relatively small compared to the CSC uncertainties, especially off-axis, and will therefore mainly affect the pairs with small combined errors. What makes it possible to separate the astrometric error from the statistical error is the fact that the former is a constant, while the CSC statistical error varies over a wide range, primarily as a function of off-axis angle.

The separation is a single-axis radial measure and, in order to perform the analysis correctly, the positional uncertainties also need to be converted to a single-axis radial quantity. CSC provides the major and minor axes of an error ellipse, while the SDSS gives independent errors in RA and Dec, which are also assumed to represent an error ellipse. However, in version 1 of the CSC the error ellipses are not fully implemented, yet, and instead approximated by circles (i.e., equal major and minor axes). The error ellipse in the SDSS is also close to a circle and thus the fact that SDSS did not report covariance for RA and Dec errors is not very important. This justifies the use of a single-axis radial combined error. We derive this radial error by adding the geometric means of the major and minor axes for CSC and SDSS in quadrature; in other words: the square root of the sum (CSC plus SDSS) of the products of major and minor axis. We want to be dealing with 1-σ values and since the CSC error ellipses refer to a 95% confidence level, the CSC values are to be multiplied by 0.408539.

To state this in a more exact fashion, we define the following quantities:

$\varepsilon_0$ : semi-major axis of CSC 95% confidence ellipse

$\varepsilon_1$ : semi-minor axis of CSC 95% confidence ellipse

$\sigma_{RA}$ : 1-σ error in RA for SDSS positions

$\sigma_{Dec}$ : 1-σ error in Dec for SDSS positions

$\sigma_c$ : 1-σ combined statistical radial position error for CSC-SDSS cross-matches

$\sigma_a$ : 1-σ astrometric error

$\sigma_c'$ : 1-σ combined corrected statistical radial position, including astrometric error

ρ : (radial) separation of CSC and SDSS positions for a cross-match pair: measured error

$\rho_N(\sigma)$ : normalized sample error

$\tilde{\chi}^2$ : reduced $\chi^2$

Then four of these quantities can be expressed as:

$$\sigma_c = \sqrt{0.1669041 \,.\, \varepsilon_0 \,.\, \varepsilon_1 + \sigma_{RA} \,.\, \sigma_{Dec}}$$

$$\sigma_c^{'} = \sqrt{0.1669041 \,.\, \varepsilon_0 \,.\, \varepsilon_1 + \sigma_{RA} \,.\, \sigma_{Dec} + \sigma_a^2}$$

$$\rho_N\,(\sigma) = \frac{\rho}{\sigma}$$

$$\tilde{\chi}^2 = \frac{\sum_1^n {\rho_N}^2}{n-1}$$

In the following, $\sigma_c$ (or $\sigma_c^{'}$) is the independent variable, $\rho$ or $\rho_N$ the dependent variable. All values are in units of arcsecond.

## 4 Analysis

After sorting the data in increasing order of $\sigma_N$ we calculated $\tilde{\chi}^2(\rho_N\,(\sigma_c))$ for bins of, successively, 100, 200, 300, 400, 500, 500, …, 500, and 476 sources, and plotted the results against the mean value of $\sigma_c$ for each bin. The result is shown in Fig. 1a. The values at $\sigma_c > 0.25$ are quite reasonable, but the steep rise below this point is indicative of an error component that is of the same order. We interpret this as caused by the missing astrometric error discussed in the Introduction. Our assertion is that, if the left hand part of the curve can be flattened out by adding a suitable value for $\sigma_a$ in $\sigma_c^{'}$ and using that value in the calculation of $\rho_N(\sigma_c^{'})$ and $\tilde{\chi}^2$ $(\rho_N\,(\sigma_c^{'}))$, one has determined the astrometric error. A value of $\sigma_a$ = 0.16″ (±0.01) provides a good result as shown in Fig.

1c. For comparison, the same plot for values of $\sigma_a$ = 0.15″ and $\sigma_a$ = 0.17″ is presented in Figs. 1b and 1d, respectively.

To verify the reliability of the result, we plotted the distribution of $\rho_N$ in three ranges of the independent variable: $\sigma_c$ < 0.15″, 0.25″ < $\sigma_c$ < 1.0″, and 1.0″ < $\sigma_c$ (Fig. 2). We expect these to show Rayleigh distributions; they do, but the one in Fig. 2a is significantly shifted toward higher values, as is to be expected. When we make the same plots again, using $\sigma_c^{'}$ and $\rho_N$ ($\sigma_c^{'}$) instead (see Fig. 3), the distributions all match. The distribution of $\rho_N$ ($\sigma_c^{'}$) for the entire sample is shown in Fig. 3d.

In Fig. 4 we present, for the bin sizes from Fig. 1, the average estimated error ρ against the average off-axis angle θ (in minutes of arc), including the 0.16 arcsec systematic error. As expected, small errors are predominantly found at small off-axis angles, large ones at large angles.

Finally, in Fig. 5 we present, for the bins from Fig. 1, the average source separation ρ against the average estimated error $\sigma_c^{'}$, including the 0.16 arcsec systematic error; the dashed black line represents the identity relation. The figure shows that ρ tracks $\sigma_c^{'}$ quite well. But the divergence at higher values of $\sigma_c^{'}$ indicates that the statistical errors of the CSC positions are likely to be overestimated when those errors are large; this corresponds (cf. Fig. 4) to off-axis angles greater than 7 or 8 arcmin. The same phenomenon can be observed in Fig. 1c where the plot slopes down for large values of the error.

We performed one more check on the results by calculating the $\tilde{\chi}^2$ function for varying ranges of off-axis angle θ. It appears that the function exhibits differences in slope on the low-error side, depending on $\theta$. In Fig. 6 we show the equivalent of Fig. 1 for all source pairs (dashed line) and for source pairs with $\theta < 5'$ (solid line) assuming an absolute astrometric error $\sigma_a = 0.18''$. However, drawing definite conclusions on the basis of this analysis is uncertain as yet, as the true nature of potential variations in positional errors across the field of view, and the associated change in PSF, is not fully understood. Consequently, there is not sufficient reason to change our recommendation of adopting $\sigma_a = 0.16''$.

## 5 Conclusion

Our conclusion is that the astrometric error in CSC positions, resulting from the four components listed in the Introduction, is 0.16″ ±0.02″. Adding this value in quadrature to the statistical error associated with each individual detection will result in a reliable value for the absolute position errors in the CSC. This will be effected for the published position errors in all releases of the Chandra Source catalog, starting with Version 1.1.

## Acknowledgments

This research has made use of data obtained from the Chandra Source Catalog, provided by the Chandra X-ray Center as part of the Chandra Data Archive. Support for the development of the Chandra Source Catalog and for this research is provided by the National Aeronautics and Space Administration through the Chandra X-ray Center, which is operated by the Smithsonian Astrophysical Observatory for and on behalf of the National Aeronautics and Space Administration under contract NAS 8-03060. The CSC Web Site is http://cxc.cfa.harvard.edu/csc/.

Funding for the SDSS and SDSS-II has been provided by the Alfred P. Sloan Foundation, the Participating Institutions, the National Science Foundation, the U.S. Department of Energy, the National Aeronautics and Space Administration, the Japanese Monbukagakusho, the Max Planck Society, and the Higher Education Funding Council for England. The SDSS Web Site is http://www.sdss.org/. The SDSS is managed by the Astrophysical Research Consortium for the Participating Institutions.

The authors wish to thank Frank Primini and Kenny Glotfelty for very helpful discussions.

## Figure captions

Fig. 1 $\tilde{\chi}^2$ ($\rho_N$ ($\sigma_c$)) and $\tilde{\chi}^2$ ($\rho_N$ ($\sigma_c'$)) as a function of $\sigma_c$, respectively $\sigma_c'$, for bins of 100, 200, 300, 400, 500, … , 500, 476 sources. (a) $\tilde{\chi}^2$ ($\rho_N$ ($\sigma_c$)) for uncorrected errors $\sigma_c$. (b) $\tilde{\chi}^2$ ($\rho_N$ ($\sigma_c'$)) for corrected errors $\sigma_c'$, where $\sigma_a$=0.15 arcsec. (c) as panel b, for $\sigma_a$=0.16 arcsec. (d) as panel b for $\sigma_a$=0.17 arcsec.

Fig. 2 The distribution of $\rho_N$ ($\sigma_c$) as a function of $\sigma_c$ . (a) for $\sigma_c$ < 0.15″; note that the Rayleigh distribution is clearly shifted to the right. (b) for 0.25″ < $\sigma_c$ < 1.0″. (c) for 1″ < $\sigma_c$.

Fig. 3 The distribution of $\rho_N$ ($\sigma_c'$) as a function of $\sigma_c'$ for $\sigma_a$ = 0.16″. Note that all Rayleigh distributions agree within reasonable limits. (a) for $\sigma_c$ < 0.15″. (b) for 0.25″ < $\sigma_c$ < 1.0″. (c) for 1″ < $\sigma_c$. (d) for all values of $\sigma_c$

Fig. 4 The measured position error (i.e., CSC-SDSS separation ρ; in arcsec) as a function of off-axis angle θ (in arcmin) averaged in bins of 100, 200, 300, 400, 500,…, 500, 476 source pairs, ordered by θ.

Fig. 5 The average value of the source separation, ρ, versus the average value of the estimated error (including a 0.16” astrometric systematic error) from the bins in Fig. 1. The dashed black line indicates the identity function. The divergence at higher values in this plot (as well as the corresponding slope in Fig. 1c) hints that the position errors at off-axis angle greater than 7-8’ (see Fig. 4) may be overestimated.

Fig. 6 $\tilde{\chi}^2$ ($\rho_N$ ($\sigma_c'$)) as a function of $\sigma_c'$, as in Fig. 1, using $\sigma_a$ = 0.18″. The dashed line represents all sources and is similar to what is presented in the panels of Fig. 1. The solid line represents only sources within an off-axis radius θ = 5□.

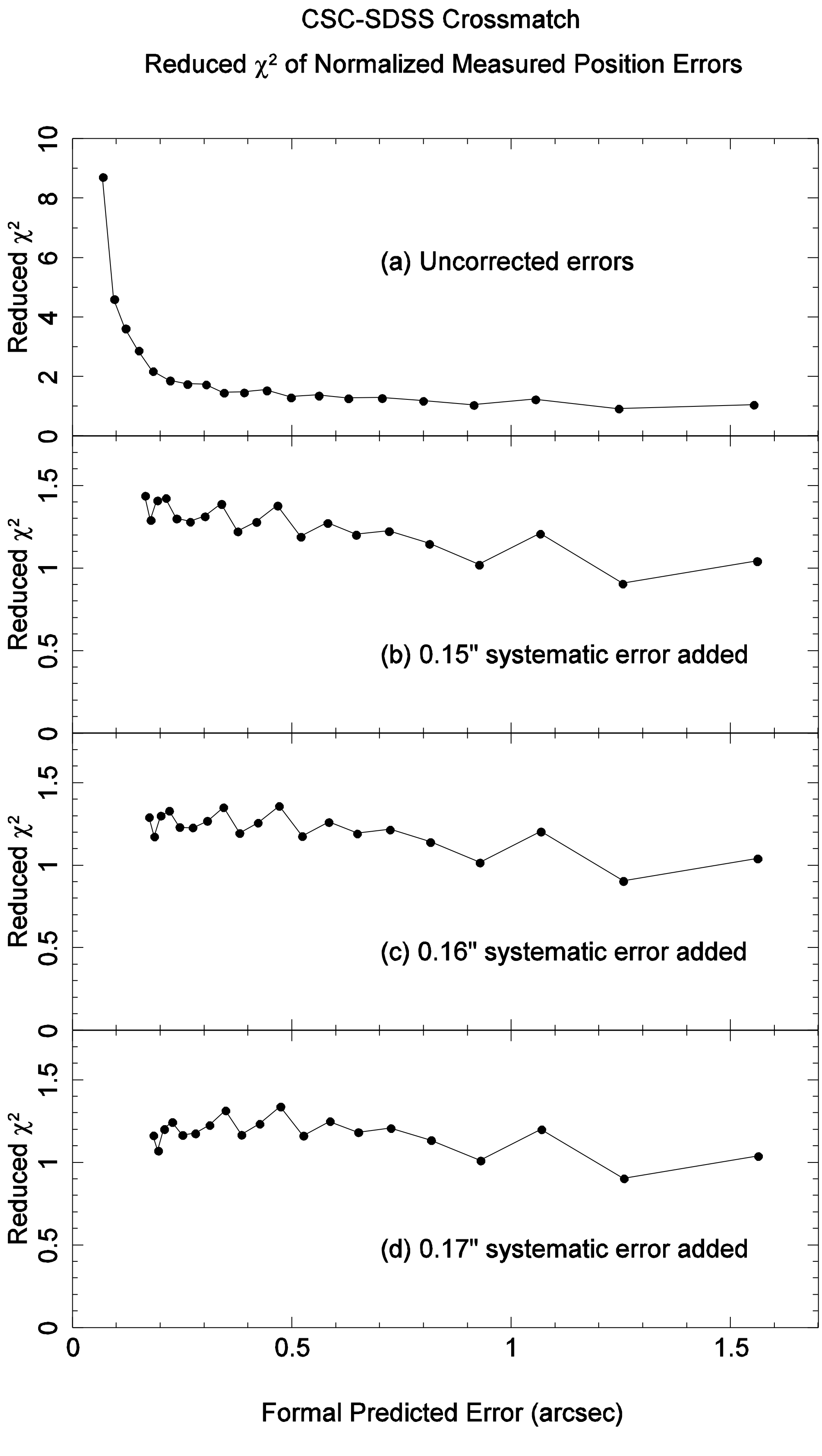
CSC-SDSS Crossmatch
Reduced χ² of Normalized Measured Position Errors
Reduced χ²
(a) Uncorrected errors
(b) 0.15" systematic error added
(c) 0.16" systematic error added
(d) 0.17" systematic error added
0
0.5
1
1.5
Formal Predicted Error (arcsec)

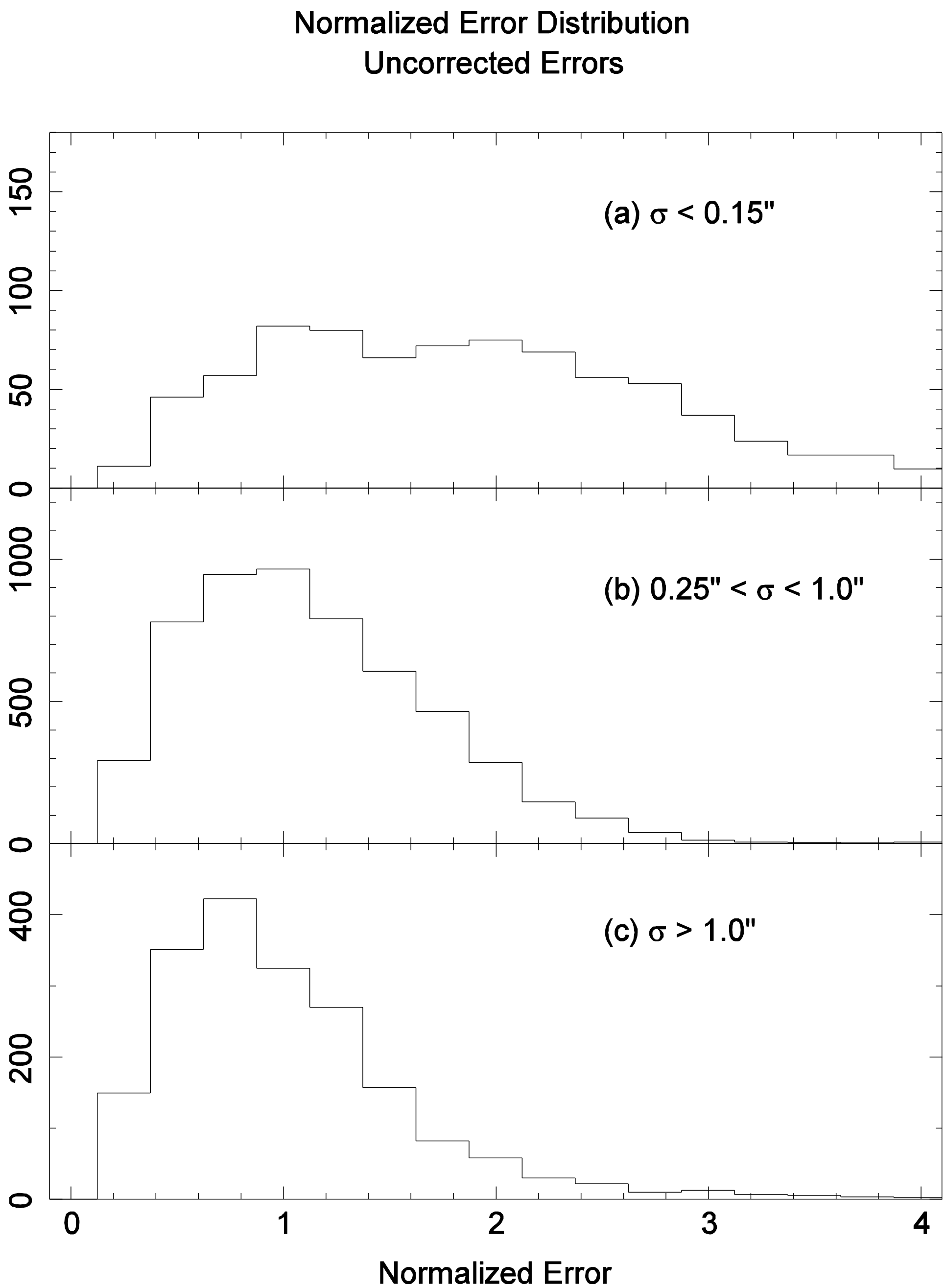
Normalized Error Distribution
Uncorrected Errors
(a) σ < 0.15"
(b) 0.25" < σ < 1.0"
(c) σ > 1.0"
0
50
100
150
0
500
1000
0
200
400
0
1
2
3
4
Normalized Error

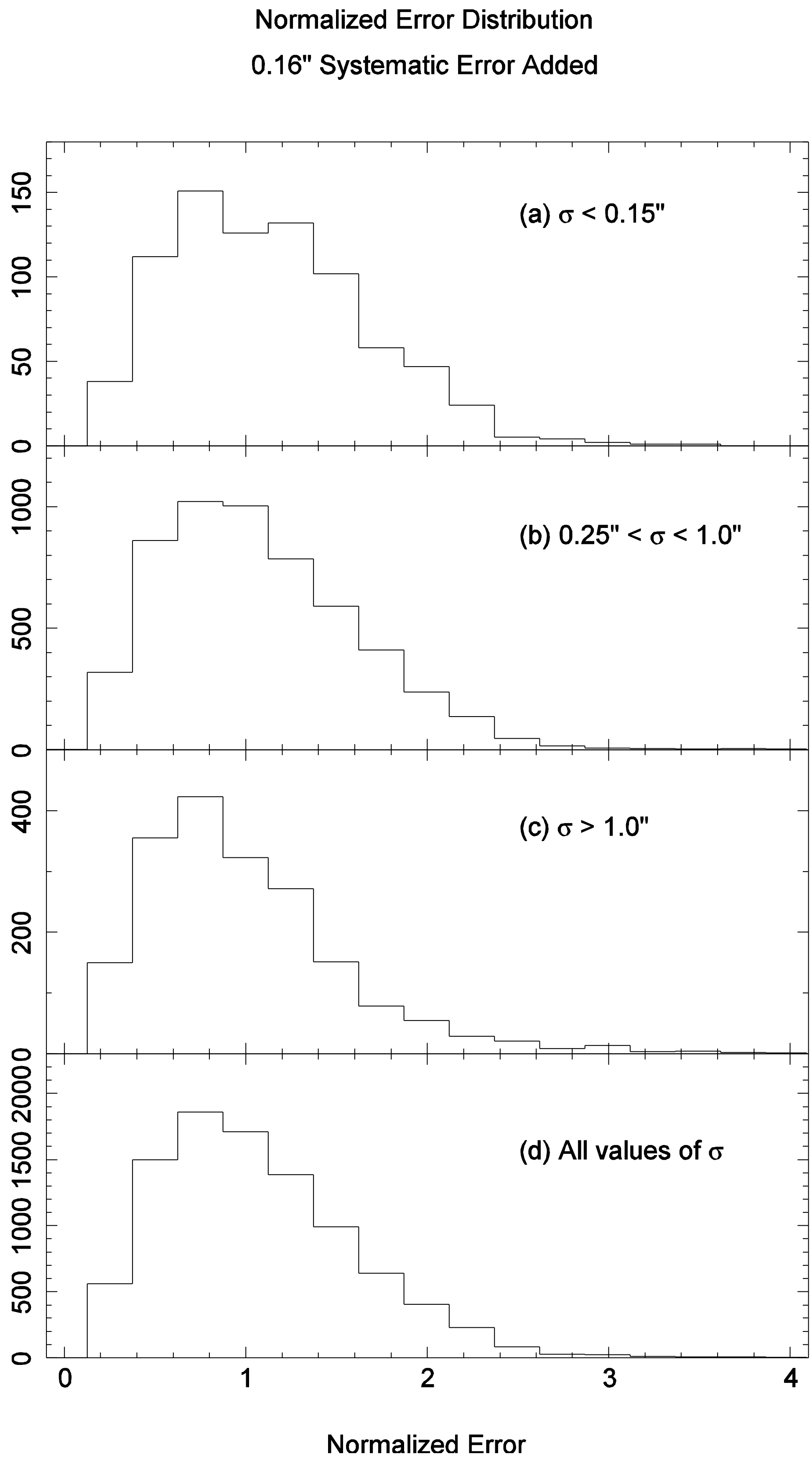
Normalized Error Distribution
0.16" Systematic Error Added
(a) σ < 0.15"
(b) 0.25" < σ < 1.0"
(c) σ > 1.0"
(d) All values of σ
0
50
100
150
0
500
1000
0
200
400
0
500
1000
1500
2000
0
1
2
3
4
Normalized Error

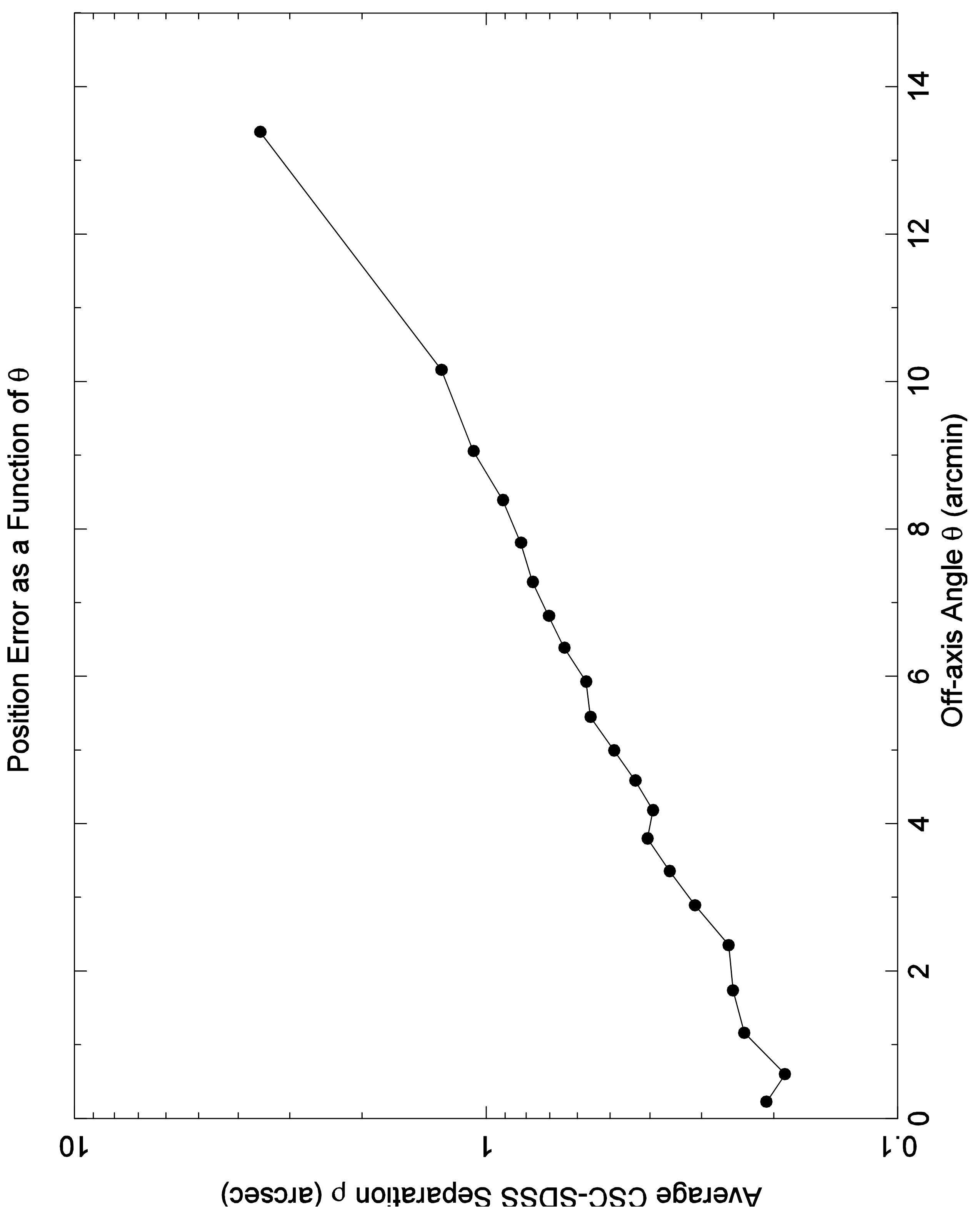
Position Error as a Function of θ
Off-axis Angle θ (arcmin)
Average CSC-SDSS Separation ρ (arcsec)
0.1
1
10
0
2
4
6
8
10
12
14

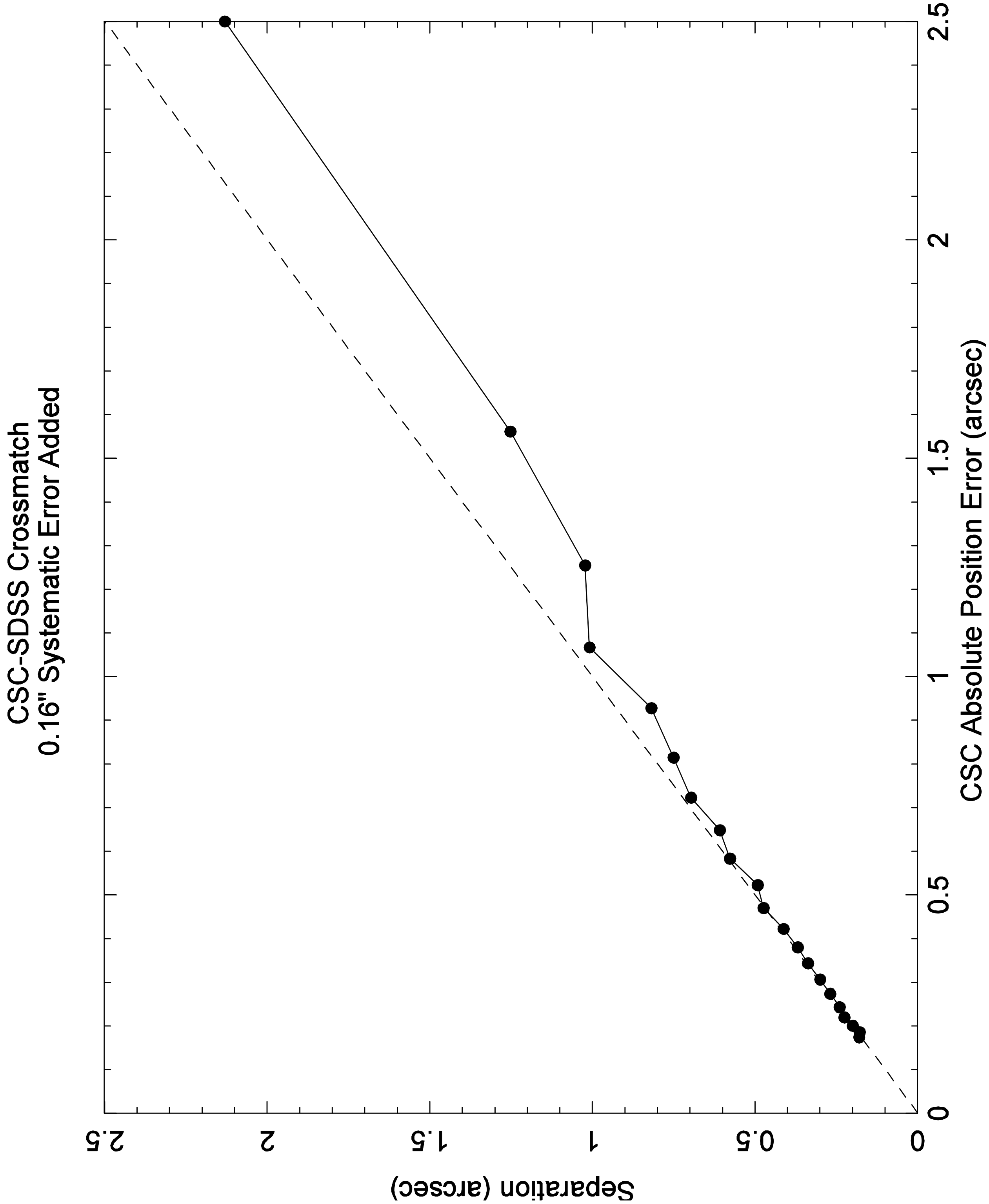
CSC-SDSS Crossmatch
0.16" Systematic Error Added
CSC Absolute Position Error (arcsec)
Separation (arcsec)
0
0.5
1
1.5
2
2.5

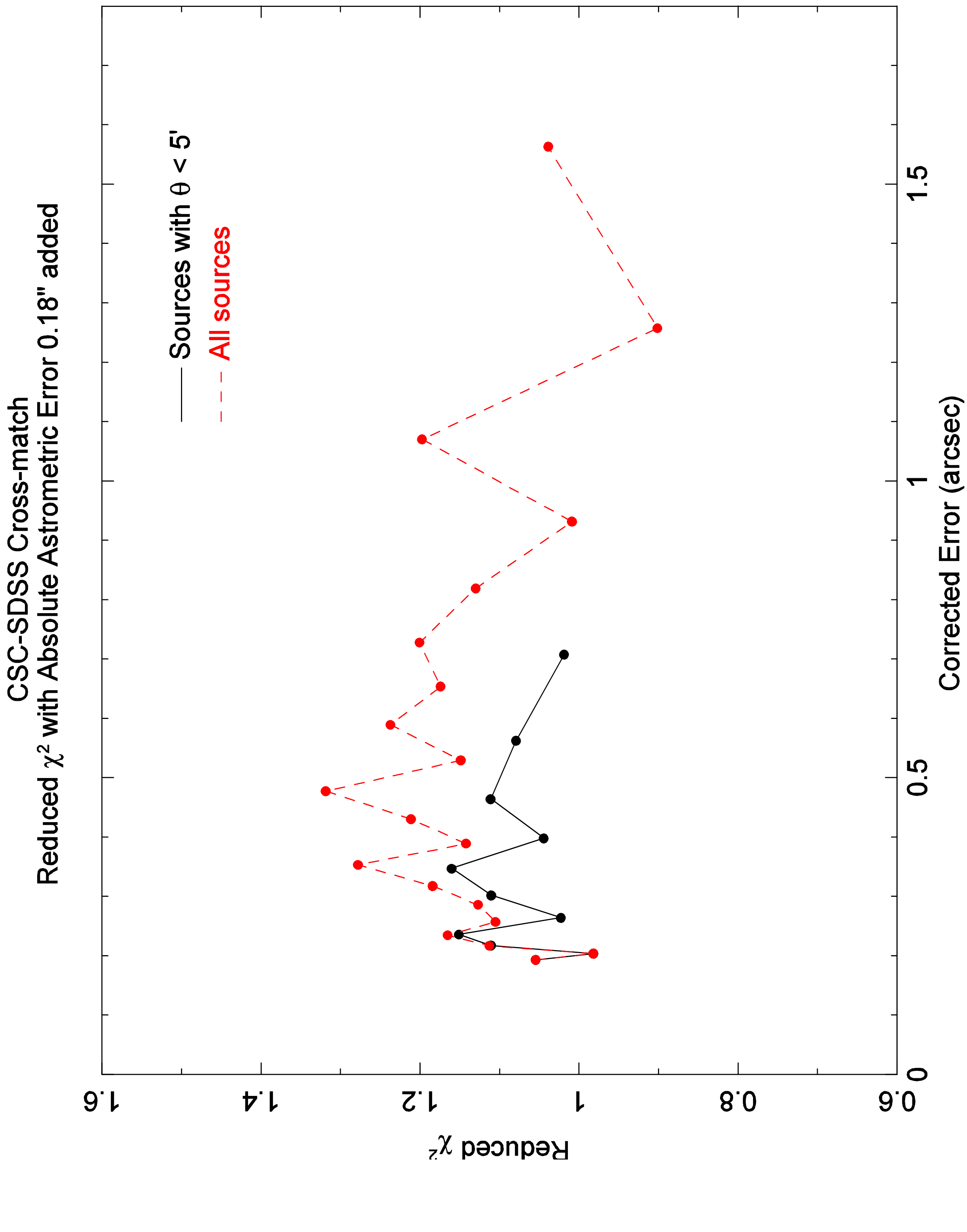
CSC-SDSS Cross-match
Reduced χ2 with Absolute Astrometric Error 0.18" added
Sources with θ < 5'
All sources
Corrected Error (arcsec)
Reduced χ2
0
0.5
1
1.5
0.6
0.8
1
1.2
1.4
1.6